\documentclass[twocolumn,showpacs,preprintnumbers,amsmath,amssymb,showpacs]{revtex4}
\usepackage{graphicx}
\usepackage{dcolumn}
\usepackage{bm}
\usepackage{amssymb}

\newcommand{\be}{\begin{equation}}
\newcommand{\ee}{\end{equation}}
\newcommand{\ba}{\begin{eqnarray}}
\newcommand{\ea}{\end{eqnarray}}

\begin{document}

\title{Perturbations of the Yang-Mills field in the universe}

\author{Wen Zhao}
\email{Wen.Zhao@astro.cf.ac.uk}\affiliation{Department of
Physics, Zhejiang University of Technology, Hangzhou, 310014,
 China}\affiliation{School of Physics
and Astronomy, Cardiff University, Cardiff, CF24 3AA, United
	Kingdom}

%\date{\today}

%%%%%%%%%%%%%%%%%%%%%%%%%%%%%%%%%%%%%%%%%%%%%%%%%%%%%%%%%%%%%%%%%%%%%%%%%%
%%%%%%%%%%%%%%%%%%%%%%%%% Abstract %%%  Abstract        %%%%%%%%%%%%%%%%%%%%%%%%%%%%%%%
%%%%%%%%%%%%%%%%%%%%%%%%%%%%%%%%%%%%%%%%%%%%%%%%%%%%%%%%%%%%%%%%%%%%%%%%%%

\small

\begin{abstract}

{\small  It has been suggested that the Yang-Mills (YM) field
can be a kind of candidate for the inflationary field at high
energy scales or the dark energy at very low energy scales,
which can naturally give the equation of state $-1<\omega<0$ or
$\omega<-1$. We discuss the zero order and first
order Einstein equations and YM field kinetic energy equations of the
free YM field models. From the zero order equations, we find that
$\omega+1\propto a^{-2}$, from which it follows that the equation of state
of YM field always goes to $-1$, independent of the initial
conditions. By solving the first order Einstein equations and YM
field equations, we find that in the YM field inflationary models,
the scale-invariant primordial perturbation power spectrum cannot
be generated. Therefore, only this kind of YM field is not enough to account for
inflationary sources. However, as a kind of candidate of dark
energy, the YM field has the `sound speed' $c_s^2=-1/3<0$, which
makes the perturbation $\phi$ have a damping behavior at the large scale.
This provides a way to distinguish the YM field dark energy models
from other kinds of models. }

\end{abstract}


\maketitle

%%%%%%%%%%%%%%%%%%%%%%%%%%%%%%%%%%%%%%%%%%%%%%%%%%%%%%%%%%%%%%%%%%%%%%%%%%%%%%%%%%%%%
%%%%%%%%%%%%%%%%%%%%%%%   SECTION 1     %%   SECTION 1  %%   SECTION 1   %%%%%%%%%%%%%%%%%%%%%%%%%%%%%%%%
%%%%%%%%%%%%%%%%%%%%%%%%%%%%%%%%%%%%%%%%%%%%%%%%%%%%%%%%%%%%%%%%%%%%%%%%%%%%%%%%%%%%%

\section{Introduction}           %% first-level sections will be auto-capitalized
\label{sect:intro}

It is well known that the physics of inflationary fields and
dark energy are two of the most important problems in
modeling cosmology. They all need a kind of matter with negative
pressure $p\simeq-\rho$ (Kolb \& Turner \cite{inf}; Bennett et al.
\cite{map}; Riess et al. \cite{sn}; Tegmark et al. \cite{sdss};
Feng et al. \cite{age}), where $p$ and $\rho$ are the pressure and
energy density of matter respectively. People always describe
them with a scalar field, which can naturally give an equation
of state of $-1\leq\omega\leq0$ (Kolb \& Turner \cite{inf}; Wetterich
\cite{de}; Bharat \& Peebles \cite{peebles}). In particular, when the
potential of the scalar field is dominant, $\omega$ will go to
$-1$. Thus, the expansion of the Universe is close to the de Sitter
expansion. Recently, the observations of the cosmic microwave
background radiation (CMB) temperature and polarization
anisotropies by the Wilkinson Microwave Anisotropy Probe (WMAP)
shows that the spectral index of the primordial density
perturbation $n_s=1.20^{+0.12}_{-0.11}$ at wavenumber
k=0.002Mpc$^{-1}$ (Bennett et al. \cite{map}). Observations of
Type Ia Supernova (SNeIa), CMB and large scale structure (LSS)
(Bennett et al. \cite{map}; Riess et al. \cite{sn}; Tegmark et al.
\cite{sdss}) also suggest that the equation of state of the dark
energy may be $\omega<-1$ (Corasaniti et al.\cite{<-1}). These are
all very difficult to obtain from single scalar field models.
So it is necessary to look for a new candidate for the
inflationary field and dark energy.

Recently, a number of authors have considered using a vector field as
the candidate of the inflationary field or dark energy (Ratra
\cite{vec}). We have advised the effective YM condensate (Zhang
\cite{Zhang}; Zhang \cite{Zhang2}; Zhao \& Zhang \cite{ymc1};
Zhang et al. \cite{ymc2}) as a kind of candidate, which can be
used to describe the inflation at high energy scales and dark
energy at very low energy scales.
%There are two major reason that prompt us to study
%this system. First the scalar models, the connection of the scalar
%field to particle physics models has not been clear so far. It
%would be the theoretically preferable if the inflation model can
%be incorporated naturally as an integral part of a sensible
%particle physics model. The second reason is that the weak energy
%condition (WEC) can not be violated by the scalar field.
In our models,  a quantum effective YM condensate is used as the
source of inflation or dark energy, instead of a scalar field. This
model has the desired interesting feature: the YM field is an
indispensable cornerstone to any particle physics model with
interactions mediated by gauge bosons, so it can be incorporated
into a sensible unified theory of particle physics. Besides, the
equation of state of this field is different from that of
general matter as well as scalar fields, and the state
$\omega<-1$ can be naturally realized.

In this paper, we shall discuss the evolution of the equation of
state of the YM field and cosmic perturbations by solving the
zero and first order Einstein equations and kinetic energy equations.
From the zero order Einstein equations, we find that the YM field
can easily give a state of homogeneity and isotropy, and it can
naturally give an equation of state $\omega<-1$ or $\omega>-1$. We
also find that $\omega+1\propto a^{-2}$ from the zero order
equations. It follows that $\omega$ naturally goes to $-1$ with
the expansion of the Universe, independent of the initial
condition. By considering the evolution of cosmic
perturbations, we investigate the first order Einstein equations
and kinetic energy equations. In the simplest condition with only
an electric field, we find that this YM field has the sound speed
$c_s^2=-1/3<0$, which is very different from scalar field
models. We also find that a scale-invariant primordial
perturbation power spectrum cannot be generated, which shows that
an electric YM field alone cannot be a candidate for the
inflationary field. However, as a candidate of dark energy, YM
field makes the cosmic fluctuation $\phi$ have a damping at
 large scales. This is helpful for answering large scale damping of
the CMB anisotropy power spectrum.

\section{The Quantum Effective Yang-Mills Field}

In the quantum effective YM field dominated Universe, the
effective YM field Lagrangian is given by (Adler \cite{L2})
 \be
 \mathcal{L}_{\rm eff}=\frac{1}{2}bF\ln|{F}/{e\kappa^2}|, \label{L}
 \ee
where $\kappa$ is the renormalization scale with dimension of
squared mass, $F\equiv -\frac{1}{2}F_{\mu\nu}^aF^{a\mu\nu}$ plays
the role of the order parameter of the YM field. The
Callan-Symanzik coefficient $b=11N/24\pi^2$ for $SU(N)$ when the
fermion's contribution is neglected. For the gauge group $SU(2)$
considered in this paper, one has $b=2\times 11/24\pi^2$. For the
case of $SU(3)$, the effective Lagrangian in Eq.(\ref{L}) leads to
a phenomenological description of the asymptotic freedom for the
quarks inside hadrons (Adler \cite{L2}). It should be noted that the
$SU(2)$ YM field is introduced here as a model for a cosmic
inflationary field or dark energy, and it may not be directly
identified as QCD gluon fields, nor the weak-electromagnetic
unification gauge fields.

An explanation can be given for the form in Eq.(\ref{L}) as an
effective Lagrangian up to a 1-loop quantum correction (Pagels \&
Tomboulis \cite{L1}; Adler \cite{L2}). A classical $SU(N)$ YM
field Lagrangian is $\mathcal{L}=\frac{1}{2g_0^2}F$, where $g_0$
is the bare coupling constant. As is known, when the 1-loop
quantum corrections are included, the bare coupling $g_0$ will be
replaced by a running $g$ as used in the following (Gross\&Wilczez
\cite{Gross}; Pagels \& Tomboulis \cite{L1}; Adler \cite{L2}),
$g_0^2\rightarrow g^2=\frac{4\times12\pi^2}{11N\ln
(k^2/k^2_0)}=\frac{2}{b\ln (k^2/k_0^2)}$, where $k$ is the
momentum transfer and $k_0$ is the energy scale. To build up an
effective theory (Pagels \& Tomboulis \cite{L1}; Adler \cite{L2}),
one may just replace the momentum transfer $k^2$ by the field
strength $F$ in the following manner, $\ln(k^2/k_0^2)\rightarrow
2\ln|F/e\kappa^2|$, yielding Eq.(\ref{L}). The expression of
$1/g^2$ is based on the renormalization group estimates. As
emphasized in references (Adler \cite{L2};  Adler \& Piran \cite{Adl}),
this estimate is formally valid whenever the running coupling
$g^2$ is small in magnitude, which is true both when
$F/\kappa^2\gg1$ (given $g^2$ is small and positive) and when
$F/\kappa^2\ll1$ (given $g^2$ is small and negative).

 The attractive features of this effective YM model include the
 gauge invariance, the Lorentz invariance, the correct trace
 anomaly, and the asymptotic freedom (Pagels \& Tomboulis \cite{L1}). With the logarithmic
 dependence on field strength, $\mathcal{L}_{\rm eff}$ has a form similar to
 the Coleman-Weinberg scalar effective potential (Coleman \& Weinberg \cite{Col}), and the
 Parker-Raval effective gravity Lagrangian (Parker \& Raval \cite{Par}). The dielectric constant
 is defined by $\epsilon=2\partial \mathcal{L}_{\rm eff}/\partial F$, and in the
 1-loop order it is given by
 \be
 \epsilon=b\ln|{F}/{\kappa^2}|. \label{eps}
 \ee
 As analyzed in (Adler \cite{L2}), the 1-loop model is a universal, leading
 semi-classical approximation. Thus, depending on whether the
 field strength $|F|\geq\kappa^2$ or $|F|\leq\kappa^2$, the YM
 condensate belongs to the family of forms whose dielectric
 constant $\epsilon$ can be positive or negative. The
 properties mentioned above are still true even if 2-loop order
 corrections are taken into account, an essential feature of the
 effective model (Adler \cite{L2}; Adler \& Piran \cite{Adl}).

 It is straightforward to extend the model to the expanding
 Robertson-Walker (R-W) spacetime. For simplicity we shall work in a
 spatially flat R-W spacetime with a metric
 $ds^2=a^2(\tau)(d\tau^2-\gamma_{ij}dx^idx^j)$,
 where we have set the speed of light $c=1$,
 $\gamma_{ij}=\delta^i_j$ denoting background space is flat, and $\tau=\int(a_0/a)dt$ is the
 conformal time.
 The dominant matter is
 assumed to be the quantum YM condensate, whose effective action
 is
 $S=\int \mathcal{L}_{eff}~a^4(\tau) ~d^{4}x$,
and the Lagrangian $\mathcal{L}_{eff}$ is defined in Eq.(\ref{L}).
By variation of $S$ with respect to $g^{\mu\nu}$, one obtains the
energy-momentum tensor,
 \be
 T_{\mu\nu}=
 -g_{\mu\nu}\frac{b}{2}F\ln|{F}/{e\kappa^2}|+\epsilon
 F^a_{\mu\sigma}F^{a\sigma}_{\nu},\label{T}
 \ee
where the energy-momentum tensor is the sum of $3$ energy-momentum
tensors of vectors, $T_{\mu\nu}=\sum_{a}T^a_{\mu\nu}$. In order to
keep the total energy-momentum tensor homogeneous and isotropic,
we assume that the gauge fields are only functions of time $t$,
and $A_{\mu}=\frac{i}{2}\sigma_aA_{\mu}^a(t)$ (Zhao\&Zhang
\cite{ymc1}). YM field tensors are defined as usual: \be
F_{\mu\nu}^a=\partial_{\mu}A_{\nu}^a-\partial_{\nu}A_{\mu}^a+\epsilon^{abc}A_{\mu}^bA_{\nu}^c.\label{define_f}\ee
This tensor can be written in the form with electric and magnetic
fields as
 \be
 F^{a\mu}_{~~\nu}=\left(
 \begin{array}{cccc}
      0 & E_1 & E_2 & E_3\\
     -E_1 & 0 & B_3 & -B_2\\
     -E_2 & -B_3 & 0 & B_1\\
     -E_3 & B_2 & -B_1 & 0
 \end{array}
 \right).
 \ee
From the definition in Eq.(\ref{define_f}), we can find that
$E_1^2=E_2^2=E_3^2$ and $B_1^2=B_2^2=B_3^2$. Thus $F$ has a simple
form $F=E^2-B^2$, where $E^2=\sum_{i=1}^3E_i^2$ and
$B^2=\sum_{i=1}^3B_i^2$. Inserting (\ref{define_f}) in (\ref{T}),
we can obtain the energy density and pressure of the YM field given by
(Zhao\&Zhang, \cite{ymc1}),
 \be
  \rho=\frac{1}{2}\epsilon(E^2+B^2)+\frac{1}{2}b (E^2-B^2),\label{T00}
 \ee
 \be
 p=\frac{1}{6}\epsilon(E^2+B^2)+\frac{1}{2}b (B^2-E^2),\label{T11}
 \ee
and
 \be
 \rho+p=\frac{2}{3}\epsilon(E^2+B^2).\label{p+rho}
 \ee
Eq.(\ref{p+rho}) follows a conclusion of this study: the weak
energy condition can be violated: $\rho+p<0$, by the effective YM
condensate matter in a family of quantum states with the negative
dielectric constant $\epsilon<0$. The whole range of allowed
values of $F$ is divided into two domains with $|F|>\kappa^2$ and
$|F|<\kappa^2$, respectively. In the domain where $|F|>\kappa^2$, one
always has $\epsilon>0$, so the WEC is still satisfied. The other
domain is $|F|<\kappa^2$ in which $\epsilon<0$, so that the WEC is
now violated.

\section{The Zero Order Equations\label{section3}}

\subsection{ The Zero Order Einstein Equations}

Let us firstly investigate the Friedmann equations, which can be
written as
 \[
 \left(\frac{\dot{a}}{a}\right)^2=\frac{8\pi G}{3}\rho,~~~~~\frac{\ddot{a}}{a}=-\frac{4\pi
 G}{3}(\rho+3p),\label{fried}
 \]
where ``dot" denotes $d/dt$. If we consider the simplest case with
only an ``electric" field, as used in the previous works (Zhang et
al. \cite{ymc2}; Zhao\&Zhang \cite{ymc1}), the Friedmann equations
can be reduced to
 \be
 \frac{d\rho}{d a^3}=-\frac{4}{3}\frac{\beta\rho}{a^3},\label{drho}
 \ee
where we have defined $\beta\equiv\epsilon/b$. If
$|\beta|\ll1$, one gets
 \be
 a^2\beta={\rm constant},
 \ee
which follows the relation of the equation of state :
 \be
 \omega+1\propto a^{-2}.
 \ee
From this relation, we find that $\omega$ will run to the critical
condition with $\omega=-1$ with the expansion of the Universe. In
the next subsectioin, we will find that from the zero order YM field
kinetic equation, we can get the same result, which means that the
YM field equation and the Einstein equations are self-rational.

\subsection{The Zero Order Yang-Mills Field Kinetic Energy Equations}

 By variation of $S$ with respect to $A_{\mu}^{a}$, one obtains the
 effective YM equations (Zhao\&Zhang \cite{ymc1})
 \be
 \partial_{\mu}(a^4\epsilon~
 F^{a\mu\nu})-f^{abc}A_{\mu}^{b}(a^4\epsilon~F^{c\mu\nu})=0,
 \label{F1}
 \ee
which can be simplified as (Zhao\&Zhang, \cite{ymc1}),
 \be
 a^2E\epsilon={\rm constant}.
 \ee
If the Universe is dominated by an ``electric" YM field, the
kinetic energy equation follows as
 \be
 a^2\beta e^{\beta/2}={\rm constant},
 \ee
When $|\beta|\ll1$, then this equation becomes
 \[
 a^2\beta(1+\beta/2)\simeq a^2\beta={\rm constant}.
 \]
which also follows as
 \be
 \omega+1\propto a^{-2}.
 \ee
This is exactly the same as that of the zero order Einstein
equations.

\section{The First Order Equations}

\subsection{The First Order Einstein Equations }

As a kind of candidate of inflationary field or dark energy, it is
very important to study the evolution of perturbations in the YM field
and cosmic fluctuations. In this section, let us consider the flat
R-W metric with the scalar perturbation in the conformal Newtonian
gauge
 \be
 ds^2=a^{2}(\tau)[(1+2\phi)d\tau^2-(1-2\psi)\gamma_{ij}dx^idx^j].\label{metric}
 \ee
The gauge-invariant metric perturbation $\psi$ is the Newtonian
potential and $\phi$ is the perturbation to the intrinsic spatial
curvature. The first order Einstein equations become (Mukhanov et al. \cite{Muk}):
 \be
 -3\mathcal{H}(\mathcal{H}\phi+\psi')+\nabla^2\psi=4\pi Ga^2\delta T_0^0,\label{EE1}
 \ee
 \be
 (\mathcal{H}\phi+\psi')_{,i}=4\pi Ga^2\delta T_i^0, \label{EE2}
 \ee
 \ba
 [(2\mathcal{H}'+\mathcal{H}^2)\phi+\mathcal{H}\phi'+\psi''+2\mathcal{H}\psi'&+&\frac{1}{2}\nabla^2D]\delta_j^i-\frac{1}{2}\gamma^{ik}D_{|kj}\nonumber\\&=&-4\pi Ga^2\delta
 T_j^i,\label{EE3}
 \ea
where $\mathcal{H}=a'/a$, $D=\phi-\psi$ and the prime denotes $d/d\tau$.
From the definition of the energy-momentum tensor (\ref{T}) and
the metric
 of (\ref{metric}), one can get the first order energy-momentum
 tensor:
 \be
 \delta T_0^0=(\epsilon-b)(B\delta B-E\delta
 E)+2\epsilon E\delta
 E+{2\phi}(\epsilon-b)B^2+\frac{B^2+E^2}{2}\delta\epsilon,\label{dT00}
 \ee
 \ba
 -\delta T_i^i&=&(\epsilon-b)(-B\delta B+E\delta
 E)-{2\phi}(\epsilon-b)B^2\nonumber\\
 &+&\epsilon[(2B_2\delta B_2+2B_3\delta B_3-2E_1\delta E_1)+4(B_2^2+B_3^2)\phi]\nonumber\\
 &+&\frac{1}{2}(E^2-B^2)\delta\epsilon+\frac{2B^2-E^2}{3}\delta\epsilon,\label{dT11}
 \ea
and others are all zero, where $E\delta E=E_1\delta E_1+E_2\delta
E_2+E_3\delta E_3$, and similar for $B\delta B$.

To obtain the gauge-invariant equations of motion for cosmological
perturbations in a Universe dominated by this kind of YM field,
we insert the general equations of the energy-momentum tensor into 
$\delta T_{\nu}^{\mu}$. First of all, from the $i-j$ $(i\neq j)$
equation it follows that we can set $\phi=\psi$, since $\delta
T_{j}^{i}=0$ $(i\neq j)$. Substituting the energy-momentum tensor
$\delta T_{\nu}^{\mu}$ into the general equations and setting
$\psi=\phi$, we find:
 \ba
 -3\mathcal{H}(\mathcal{H}\phi&+&\phi')+\nabla^2\phi
 \nonumber\\
 &=&4\pi Ga^2\{(\epsilon-b)(B\delta B-E\delta E)
 +2\epsilon E\delta E
 \nonumber\\
 &+&{2\phi}(\epsilon-b)B^2+\frac{B^2+E^2}{2}\delta\epsilon\},\label{29}
 \ea
 \be
 (\mathcal{H}\phi+\phi')_{,i}=0,\label{30}
 \ee
 \ba
 [(2\mathcal{H}'&+&\mathcal{H}^2)\phi+\mathcal{H}\phi'+\phi''+2\mathcal{H}\phi']
 \nonumber\\
 &=&4\pi Ga^2\{(-B\delta B+
 E\delta E)(\epsilon-b)-{2\phi B^2}(\epsilon-b)
\nonumber\\
 &+&\frac{\epsilon}{3}[4B\delta B-2E\delta E+8\phi B^2]
 +\frac{E^2+B^2}{6}\delta\epsilon\}.\label{31}
 \ea
where $\delta\epsilon=(2E\delta E-2B\delta B-4B^2\phi)/(E^2-B^2)$,
the relation between $\delta E$ and $\delta B$ depends on the YM
field equations as below.

\subsection{The First order kinetic equations of the YM field}

The metric as before with $\psi=\phi$ is
 \be
 ds^2=a^2(\tau)[(1+2\phi)dt^2-(1-2\phi)\gamma_{ij}dx^idx^j],
 \ee
then $\sqrt{-g}=a^4(1-2\phi)$. The equation kinetic energy equation is
 \be
 \partial_{\mu}(a^4(1-2\phi)\epsilon~
 F^{a\mu\nu})-f^{abc}A_{\mu}^{b}(a^4(1-2\phi)\epsilon~F^{c\mu\nu})=0,
 \ee
from which, one can get the first order perturbation equations:
 \be
 \partial_{\mu}[a^2(\epsilon\partial_{\tau}\delta
 A+\delta\epsilon\partial_{\tau}A-2\epsilon\phi\partial_{\tau}A)]=0,~~~(\mu=0,1,2,3)
 \ee
 \be
 \partial_{i}[\delta\epsilon A^2+2\epsilon\phi A^2+2\epsilon A\delta A]=0,~~~(i=1,2,3)
 \ee
 which also can be written as
 \be
 \partial_{i}(B\delta\epsilon+2\epsilon\phi B+2\epsilon \delta
 B)=0,
 \ee
 \be
 \partial_{i}(\epsilon\delta E+\delta \epsilon E-2\epsilon\phi
 E)=0,
 \ee
 \be
 \partial_{\tau}[a^2(\epsilon\delta E+\delta\epsilon E-2\epsilon\phi
 E)]=0.
 \ee
 From these equations, we can immediately get the simple relation between
 $\delta B$ and $\delta E$:
 \be
 a^2(\epsilon\delta E+\delta\epsilon E-2\epsilon\phi
 E)={\rm constant}.
 \ee
 which is useful when solving the first order Einstein equations (\ref{29})-(\ref{31}).

\subsection{The Solution of the Perturbations. }

Here we only discuss the simplest case with $B\equiv 0$, the first
order energy-momentum tensor becomes very simply
 \be
 \delta T_0^0=(\epsilon+2b)E\delta
 E,\label{dT00}
 \ee
 \be
 -\delta T_i^i=(\epsilon-2b)E\delta
 E/3,\label{dT11}
 \ee
 \be
 \delta T_i^0=-2\epsilon[E_1(\delta B_3-\delta B_2)]=0,\label{dT01}
 \ee
 where we have used  $\delta \epsilon=2bE\delta E/E^2$ when $B\equiv0$.
 From Eq.(\ref{dT00})-(\ref{dT01}), we find
 that $\delta T^{\mu}_{\nu}$ is independent of the metric perturbation of
 $\phi$, which is different from the scalar field models
 (Weller \& Lewis \cite{Wel}; Armendariz-Picon  et al. \cite{Arm}; DeDeo et al. \cite{Deo}). The first order Einstein
 equations become
%  so in the frame in which the YM field
% is unperturbed $\delta E=0$, then $\delta\rho=\delta p\equiv0$, the "dark
% energy" is an exact de Sitter background if $\epsilon=0$, which is difficult to
% distinguished with the $\Lambda CDM$ cosmic models,
% and this model fit very well with the CMB observation\cite{map1}\cite{map2}\cite{Efs}.
% If we consider the perturbation  $\delta E$, from the first order Einstein equations, we can get
 \be
 -3\mathcal{H}(\mathcal{H}\phi+\phi')+\nabla^2\phi=4\pi Ga^2E\delta
 E(\epsilon+2b),\label{EEE1}
 \ee
 \be
 (\mathcal{H}\phi+\phi')_{,i}=0,\label{EEE2}
 \ee
 \be
 [(2\mathcal{H}'+\mathcal{H}^2)\phi+\phi''+3\mathcal{H}\phi']=4\pi Ga^2
 E\delta E(\epsilon-2b)/3.\label{EEE3}
 \ee
 From this we obtain the main equation, which describes the evolution of the
 metric perturbation $\phi$ with time:
 \be
 \phi''+3\mathcal{H}(1-\gamma)\phi'+\gamma\nabla^2\phi+(2\mathcal{H}'+\mathcal{H}^2-3\mathcal{H}^2\gamma)\phi=0,\label{phi}
 \ee
 where $\gamma\equiv\frac{2b-\epsilon}{6b+3\epsilon}$. From this equation,
 one can easily find that the evolution of $\phi$ only depends on
 $\gamma$ and $\mathcal{H}$, but not on the first order YM field kinetic
 equations. This is because we have only considerd the ``electric" YM field.
 If we also consider the $B$ components, the YM equation (38)
 will be used to relate $\delta E$ and $\delta B$.

 For the inflationary field, the most important prediction is that the
 inflation can generate a scale-invariant primordial scalar perturbation power spectrum, which
 has been found in the CMB power spectrum and large scale
 structure. Now we shall firstly consider whether a Universe dominated by this YM field
 can also generate a scale-invariant spectrum using the general scalar inflationary
 models.

 If we consider the YM field as a kind of candidate of
 inflationary field, the YM field should satisfy the following
 constraints:
 1) Firstly, we should require that the inflation can exist, which
 means that the YM field has a state with $\omega<-1/3$. Using the $p$ and
 $\rho$ in Eq.(12), one can easily get a
 constraint on the YM field:
 \be
 \epsilon<b.
 \ee
 2) The energy density of the YM field should be positive, which
 induces the second constraint on the YM field:
 \be
 \epsilon>-b.
 \ee
 3) Due to equation (\ref{phi}), if we can define an
 adiabatic vacuum state at the very high frequency
 $(k\rightarrow\infty)$, which requires that $\gamma<0$, this would require the third constraint on the YM field:
 \be
 \epsilon>2b,~~{\rm or}~~\epsilon<-2b.
 \ee

 We find these three simple constraints on the YM field cannot be
 satisfied at the same time, so this model cannot generate a
 scale-invariant primordial power spectrum as a general scalar field
 inflationary model. However, it is necessary to notice
 that this does not mean that the YM field cannot be the source of
 the inflation. Recently, some authors have
 discussed a kind of curvaton reheating mechanism
 in non-oscillatory inflationary models (Feng \& Li \cite{cur}). In this kind of model,
 the primordial spectrum and the reheating can be generated by the other
 curvaton field. So, although the YM field cannot generate a scale-invariant primordial
 spectrum, the YM field can also be the background field in the
 curvaton field inflationary models, which will be discussed in a future work.

 However, the YM field can be a good candidate of dark energy (Zhao \& Zhang \cite{ymc1}; Zhang et al. \cite{ymc2}).
 Now let us discuss the evolution of the cosmic scalar fluctuations $\phi$
 in the YM field dark energy models.

 Neglecting anisotropic stress, the potential $\phi$ evolves as
 {Weller \& Lewis \cite{Wel}; Ma \& Bertschinger \cite{Ma}; Gorden \& Hu \cite{Chr})
 \ba
 \phi''+3\mathcal{H}\left(1+\frac{p'}{\rho'}\right)\phi'&-&\frac{p'}{\rho'}\nabla^2\phi
 +\left[\left(1+3\frac{p'}{\rho'}\right)\mathcal{H}^2+2\mathcal{H}'\right]\phi\nonumber\\&=&4\pi
 Ga^2\left(\delta p-\frac{p'}{\rho'}\delta\rho\right),\label{de1}
 \ea
 where $p=\sum_ip_i$ and $\rho=\sum_i\rho_i$, which should include the
 contributions of the baryon, photon, neutron, cold dark matter, and the dark
 energy. Here we consider the simplest case with only the YM field dark
 energy, and it has the equation of state$\omega_{de}=-1$. The `sound
 speed'
 is defined by
 $c_s^2=\delta p/\delta \rho$. From the Eqs. (39) and (40), one finds that
 $c_s^2=-1/3$. So then, the equation (\ref{de1}) becomes:
 \be
 \phi''+2\mathcal{H}\phi'+\frac{1}{3}\nabla^2\phi+2\mathcal{H}'\phi=0,\label{phi2}
 \ee
 which is the same as equation (\ref{phi}) with $\gamma=1/3$.
 Defining $u\equiv a\phi$, one gets
 \be
 u_k''-\frac{k^2}{3}u_k=0,
 \ee
 where $u_k$ is the Fourier component with wavenumber $k$,
 we have used $a\propto 1/\tau$, which has the solution of
 \be
 u_k\propto e^{\pm\frac{ k\tau}{\sqrt{3}}}.
 \ee
 For the large wavelength case ($k\tau\ll1$), we have
 \be
 \phi\propto a^{-1}
 \ee
 and when $k\tau\gg1$, one has the growing solution
 \be
 u\propto e^{-\frac{k\tau}{\sqrt{3}}},~~{\rm and}~~\phi \sim
 a^{-1}e^{-\frac{k\tau}{\sqrt{3}}}.
 \ee

 We find in this universe, the evolution of the cosmic fluctuation $\phi$ is very
 different from that in the scalar field dark energy models. For the fluctuation with
 large wavelength $k\tau\ll1$, it is always damped
 with the expansion of the Universe. However for the fluctuation with
 small wavelength, it has a
 rapid growth with time, which is because of the negative sound speed $c_s^2=-1/3<0$.
 This should have an important effect on the
 large scale integrated Sachs-Wolfe effect of CMB.
 We should mention that the result in this subsection is only qualitative since we have not considered other
 effects, such as the baryons, photons, neutrons, dark matter and so on,
 which are also important for the evolution of the cosmic fluctuation $\phi$, especially at small scales.

\section{Conclusion and Discussion }

In this paper, we have investigated the effective YM field as a
kind of candidate for an inflationary field or dark energy. From the zero
order Einstein equations and YM field kinetic energy equations, we find
that this field can naturally give an equation of state
$-1<\omega<0$ and $\omega<-1$. This is one of the most important
feature of this model. Also, we get $\omega+1\propto a^{-2}$,
which suggests that $\omega$ will go to $-1$ with the expansion
of the Universe. This makes the Universe dominated by this field
have a de Sitter expansion.

When considering the evolution of the perturbations, we solved the
first order Einstein equations and YM field kinetic energy equations. We
find that the model with only an electric field cannot generate a
scale-invariant primordial scalar perturbation power spectrum.
This means that a model that only used the YM field is not a good candidate for describing the
inflationary field. However, as a kind of candidate of dark
energy, we obtained the equation of the YM perturbation, and found
that this field is very different from the general scalar field
dark energy models. This YM field has a negative sound speed,
which makes the cosmic perturbations dampen at large scales.
This is a very important source of the integrated Sachs-Wolfe
effect of the CMB power spectrum. This damping of $\phi$ at large
scale may be helpful for answering the very small quadrupole problem
of the CMB temperature anisotropy power spectrum.

We should mention that, in this paper we have not considered the
possible interaction between the YM field dark energy and the
other components, especially the dark matter (Zhang et al. \cite{ymc2}).
We leave this topic to future work.

~

{\bf Acknowledgement:}

W.Zhao appreciates useful discussions with Prof.Y.Zhang. This work
is supported by Chinese NSF grants No.10703005 and No.10775119.

%%%%%%%%%%%%%%%%%%%%%%%%%%%%%%%%%%%%%%%%%%%%%%%%%%%%%%%%%%%%%%%%%%%%%%%%%%
%%%%%%%%%%%%%%%%%%%%%%%%% APPENDICES %%  Appendix   %%%%%%%%%%%%%%%%%%%%%%%%%%%%%%
%%%%%%%%%%%%%%%%%%%%%%%%%%%%%%%%%%%%%%%%%%%%%%%%%%%%%%%%%%%%%%%%%%%%%%%%%%

%%%%%%%%%%%%%%%%%%%%%%%%%%%%%%%%%%%%%%%%%%%%%%%%%%%%%%%%%%%%%%%%%%%%%%%%%%%%%%%%%%%%%
%%%%%%%%%%%%%%%%%%%      References    References   References    %%%%%%%%%%%%%%%%%%%%%%%%%%%%%%%%%%%%%%
%%%%%%%%%%%%%%%%%%%%%%%%%%%%%%%%%%%%%%%%%%%%%%%%%%%%%%%%%%%%%%%%%%%%%%%%%%%%%%%%%%%%%

\end{document}